# Studying real-world societal problems in a STEM context through robotics

**Chronis Kynigos**

National and Kapodistrian University of Athens

Educational Technology Lab

kynigos@ppp.uoa.gr

**Marianthi Grizioti**

National and Kapodistrian University of Athens

Educational Technology Lab

mgriziot@ppp.uoa.gr

**Christina Gkreka**

National and Kapodistrian University of Athens

Educational Technology Lab

xristgreka@gmail.com

## Abstract

In this paper, we will discuss the design and implementation of three educational robotics workshops that focus on the use of robotics for addressing real-world societal problems. Through the workshops, we aimed to investigate how the integration of real-world problems to robotics can provide a multidisciplinary approach for 'child-robot interaction' that allows students to engage with different STE(A)M concepts. In addition, we examined if this integration increased or strengthened students' interest in STEM/robotics education and careers. The workshops were designed and implemented in the context of the European project Educational Robotics for STEM (ER4STEM) with students from different age groups using Lego Mindstorms and Arduino kits.

## ACM Classification Keywords

I.2.9 Robotics; K.3.2 [Computers and Education]: Computer and Information Science Education;

## Introduction

During the last two decades, activities with robots have progressively found their way in formal educational

**The ER4STEM Project**

- Robotics workshops and competitions in 7 European Countries
- >4000 students aged 7-19
- Robotics kits include: Lego Mindstorms, Lego WeDo, Thymio, Arduino, BiBot etc.

Data Collection:

- interviews, pre and post questionnaires, video recordings with the consent of students and parents, reflection documents and the programs and robotic artefacts of the students

Data analysis:

- qual + quant mixed methods
- in-depth case studies
- cross-country analysis

systems. In parallel, in the last few years there has been a big effort worldwide, for the inclusion of all young learners in STEM (Science, Technology, Engineering, and Mathematics) education and for their attraction to STEM-related careers. However, regarding STEM education, educational robotics activities often focus either on the 'Technology' (programming) or the 'Engineering' (constructing) element of STEM, with the 'Science' and 'Mathematics' lying somewhere on the background. In addition, there is a lack of robotics activities which connect robotics and STEM to real-world situations by, for example, implementing robotic solutions for common societal issues. Most of the approaches do not empower children to define problems that influence their lives and provide them with the necessary skills to solve these. However, this connection is quite important for the attraction of more students both to STEM education and also to STEM and robotics careers. We believe that in order to gain and sustain students' interest to STEM, their interaction with robotics should occur in an authentic and multidisciplinary context which provides a view of how robotics are used in real-world situations. In this paper, we will discuss three workshops that engaged students in the discussion, design, construction and presentation of robotic constructions for solving real-world problems. The workshops' activities integrated elements from the production process and resemble the use of robotics in realistic situations. The robotic activities based on the constructionist learning approach [2]. Thus they supported meaning generation and exchange of ideas through collaboration, experimentation and construction. Our aim was to investigate how the connection of robotics to real-world problems through open, constructionist activities can provide an authentic and multidisciplinary context for future robotics activities. The workshops were designed and implemented as part of the European project ER4STEM, a three-year project funded by the European Commission that focuses on the development of an integrated framework for the different approaches in using robotics in education [3].

**Real-world Societal Problems in Educational Robotics**

One of the ER4STEM targets is to study real-world societal problems as perceived by each child and relate societal challenges to existing technologies. Thus, we implemented a number of workshops that focus on recognizing and addressing such issues with the use of robotics. The workshops were designed with the "activity plan template", a design tool for planning robotics activities that depicts what we have identified as essential and transferrable elements of learning with robotics [4]. An activity plan for a workshop, apart from the sequence and description of activities, it also contains information about the focused STEM concepts and learning skills, the expected outcomes, the collaboration methods etc. The workshops presented in this paper, involved students in a real-world problematic situation, through three important processes for constructionist learning with robots: Construction, Programming and Argumentation. While students are switching between these three processes, they express their personal perceptions and ideas on the problem and discuss suggested solutions. They also apply knowledge from different STEM fields in an authentic context in order to create personally meaningful objects and address real-world societal needs. Through this process, we aim to support students to connect robots to their personal interests and share their ideas with these tangible artefacts while learning basic scientific concepts.

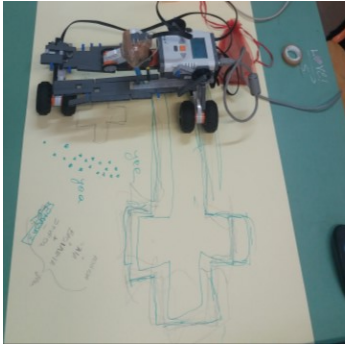

Figure 1: The CNC robotic machine draws a cube net

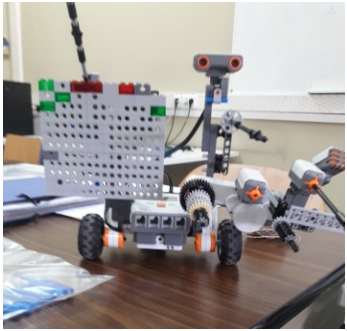

Figure 2: A robotic vehicle that would help persons with impaired mobility to park

**The three workshops**

In the above context, we will discuss the design, implementation and preliminary results of three educational robotics workshop organized with different age groups. All of them engaged students with a societal issue which they had to address, with the use of robotics. For the evaluation of these workshops we focused on evidence of students' engagement with STEM concepts (all data), discussion on societal issues (videos), change of interest to STEM or STEM careers (questionnaires, interviews), development of the necessary skills to address such issue such as collaboration, creativity, communication etc (all data).

The first workshop deals with the emergency situation of housing a large number of people (i.e. due to migration or a physical disaster). Students are asked to create physical models of temporary houses, simulating the practices of professionals who design and build containers. Firstly they design virtual models of houses using the digital tool "MaLT" (etl.ppp.uoa.gr/malt2) that supports the programming of 3D dynamic geometrical models. Then they build a robotic 'cutter' with Lego Mindstorms that resembles a CNC (Computer Numerical Control) milling machine; a machine that cuts solids' nets. Finally, they program their robot to draw cubes and other solids' nets on the paper (Figure 1). These drawings are cut and tiled in order to create physical models of container houses which are placed on a maquette. This workshop involves students in the production process of a final product which can be used in a real situation. Its activities enclose concepts from mathematics (geometrical nets, 3D geometry, angles, scale etc), engineering (gears, movement in different axes), computer science (programming of the robot), architecture (modelling, building design). In the workshop participated 18 students from a primary school aged 10-11 years old and it was implemented in the school context, as part of the computer science subject.

The second workshop engages students with the issue of inclusion of persons with impaired mobility. First students are given a story about one person with impaired mobility who tried to park his car but all the parking spots for persons with special needs were taken. Then they discuss the story and propose ideas for a robotic vehicle that would help that person. Each team constructs their 'smart' vehicle using Lego Mindstorms and extra sensors (distance sensor, color sensor, ultrasonic sensor) and program it to move smoothly, avoid obstacles, park between to objects etc. In the last phase, each team presents their final vehicle to an imaginary car company. In this phase, students have to describe the functionalities of their robots, calculate an estimate construction cost but also highlight its benefits for the society. The workshop engages students with different STEM fields including mathematics (angle, distances calculation, rotation), physics (vehicle movements, friction), engineering (distance sensor, vehicle construction) and computer science (loops, conditionals, events). The activities were implemented as a project of the school's mathematics club and participated 24 students in total, aged 13-15 years old.

The third workshop was an extension of an older activity for engaging high school students with STEM concepts through electronics [2]. Students designed and constructed a robotic insect with the combination of Arduino, electronic sensors and everyday material such as tape, ropes and paper clips. Their insect would be used for addressing an environmental issue of their choice. For example, one team decided to build a robotic scorpion with eyes and sensors in order to observe endangered scorpion species without disturbing their natural environment (Figure 3). This workshop included concepts from the fields of Electronics (Arduino circuits, sensors), Computer Science (Programming), Biology (Insect skeleton and movement) and Physics (movement, friction). It was organized as an after-school activity in a vocational school and the participants were 15 students aged 17-19 years old.

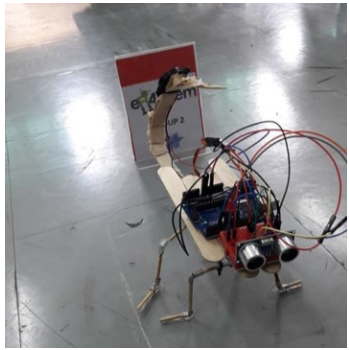

Figure 3: A robotic insect for addressing environmental issues

**Preliminary Results**

Through the analysis of the collected data we aim to answer the following questions:

- How the connection of robotics to real-world problems can increase students' attitudes and interest to STEM/robotics education and careers?
- What powerful ideas students shared and exchanged about the use of robotics in real-world situations and about STEM concepts?

The preliminary results indicate that students' applied and generated knowledge for multiple STEM concepts while they were discussing and solving the different societal issues. For instance, in the second workshop two teams used the Pythagorean Theorem in order to program the robotic vehicle to park correctly. In addition, in order to achieve a smooth and safe movement of the vehicle, they combined concepts from physics (i.e. friction) with mathematical concepts (i.e. wheel radius) and programming concepts (i.e. loops). Furthermore, in the first and the third workshop there is also evidence for increasing students' interest to STEM. For example, in the pre-questionnaires of the first workshop there were six girls who gave (very) negative answers in the regarding their opinion for science, math and ICT subjects (i.e. less favorite subject, very difficult etc), which then changed to positive answers in their post-questionnaires after the workshop. In addition four of the girls also changed their answer to the question 'What job would you like to do?' to a STEM-related one. Finally, many teams extended their initial ideas and implemented unexpected solutions for the initial problem, which indicates that the engagement with realistic problems may urge their creativity with robotics.

**References**
1. Alimisis, D. & Kynigos, C. (2009). Constructionism and robotics in education. Teacher Education on *Robotic-Enhanced Constructivist Pedagogical Methods*, (pp. 11-26).
2. Kynigos, C., Grizioti M. & Nikitopoulou S. (2017). "RobIn: A Half-baked Robot for Electronics in a STEM Context." *Proceedings of the 2017 Conference on Interaction Design and Children*. ACM, 2017.
3. Lammer, L., Lepuschitz, W., Kynigos, C., Giuliano, A., & Girvan, C. (2017). ER4STEM Educational Robotics for Science, Technology, Engineering and Mathematics. *Robotics in Education (*pp. 95-101). Springer International Publishing.
4. Yiannoutsou, N., Nikitopoulou, S., Kynigos, C., Gueorguiev, I., & Angel-Fernandez, J. (2017). Activity plan template: a mediating tool for supporting learning design with robotics. In *Robotics in Education* (pp. 3-13). Springer, Cham.